%% file: main.tex
\RequirePackage{silence}
\RequirePackage{amsmath}
\PassOptionsToPackage{pdfdisplaydoctitle={true}, pdfpagemode=UseNone, pdfauthor={Mihai Herda, Michael Kirsten, Etienne Brunner, Joana Plewnia, Ulla Scheler, Chiara Staudenmaier, Benedikt Wagner, Pascal Zwick, Bernhard Beckert}, pdftitle={Understanding Counterexamples for Relational Properties with DIbugger}, pdfsubject={Debugging}, pdfkeywords={Relational properties, software verification, debugging}, pdfcreator=pdflatex, pdfstartview={FitH}, unicode={true}, breaklinks={true}, hyperindex={true}, pdfpagelabels={true}, plainpages={false}}{hyperref}
\PassOptionsToPackage{capitalize,noabbrev,nameinlink}{cleveref}

\documentclass[submission,copyright,creativecommons]{eptcs}
\providecommand{\event}{PERR 2019}
\usepackage{underscore}

\usepackage[english]{babel}
\usepackage[utf8x]{inputenc}
\usepackage[T1]{fontenc}
\usepackage[usenames]{xcolor}
\usepackage{listings}
\usepackage[inline]{enumitem}
\usepackage{chngcntr}
\usepackage{csquotes}
\usepackage{upquote}
\usepackage{siunitx}

\usepackage{graphicx}

\usepackage[normalem]{ulem}
\usepackage{stmaryrd}
\usepackage{cite}
\usepackage{hyperref}
\usepackage{cleveref}

\newcommand{\KeY}{\texorpdfstring{Ke\kern-0.1emY}{KeY}}

\title{\texorpdfstring{Understanding Counterexamples for \\ Relational Properties with \emph{DIbugger}}{Understanding Counterexamples for Relational Properties with \emph{DIbugger}}}

\author{\qquad Mihai Herda$^*$ \qquad\quad Michael Kirsten$^\dag$ \qquad\quad Etienne Brunner \\[0.5em]
\qquad\quad Joana Plewnia \qquad\qquad Ulla Scheler \qquad\quad\; Chiara Staudenmaier \\[0.5em]
\qquad Benedikt Wagner \qquad\quad Pascal Zwick \qquad\quad\; Bernhard Beckert$^\ddag$
\institute{Karlsruhe Institute of Technology (KIT), Karlsruhe, Germany}
\email{\quad $^*$herda@kit.edu \qquad\qquad $^\dag$kirsten@kit.edu \qquad\qquad $^\ddag$beckert@kit.edu}
}
\def\titlerunning{Understanding Counterexamples for Relational Properties with \emph{DIbugger}}
\def\authorrunning{M. Herda et al.}

\begin{document}
\maketitle

\begin{abstract}
Software verification is a tedious process that involves the analysis of multiple failed verification attempts, and adjustments of the program or specification.
This is especially the case for complex requirements, e.g., regarding security or fairness, when one needs to compare multiple related runs of the same software.
Verification tools often provide counterexamples consisting of program inputs when a proof attempt fails, however it is often not clear why the reported counterexample leads to a violation of the checked property.
In this paper, we enhance this aspect of the software verification process by providing \emph{DIbugger}, a tool for analyzing counterexamples of relational properties, allowing the user to debug multiple related programs simultaneously.
\end{abstract}

\input{introduction}
\input{tool}
\input{examples}
\input{related}
\input{conclusion}

\bibliographystyle{eptcs}
\bibliography{bib}
\end{document}

%% file: introduction.tex
\section{Introduction}
\label{sec:introduction}

Software verification is a tedious process that involves the analysis of multiple failed verification attempts, and adjustments of the program or specification.
Oftentimes, this is an incremental process, where at first neither the formal specification captures the informally-given requirements, nor the program adheres to the specification.
The task becomes even trickier when the requirements are complex, as is often the case for security (e.g., noninterference for information flow~\cite{SchebenGreiner2016}) or fairness (e.g., for resource allocation~\cite{lanKCS2010} or voting~\cite{BeckertBormerKirstenEtAl2016}) requirements, which can only be captured using \emph{relational properties}.

Relational properties refer to at least two program runs.
A classical example of such a property is \emph{program equivalence}---the property that two programs provided with identical inputs generate identical outputs.
Relational properties are highly relevant in the field of evolving safety-critical systems, e.g., when modifying the software, or when one software component is replaced with another one.
When applying such a change to the software, we wish to make sure that this change does not introduce new bugs.
Relational verification tools such as LLR\^{e}ve~\cite{KieferKlebanovUlbrich2017} can prove that a new---but similar---software program is equivalent (modulo some allowed changes) to the preceding existing software.

In case the verification of these properties fails, existing verification tools can provide counterexamples.
Such counterexamples contain concrete inputs which are identical between the two programs, but for which the execution of the two programs leads to two different outputs.
Understanding \emph{why} the provided inputs are a counterexample is---however---usually not a trivial task.
Whereas this task is already difficult for functional properties, it becomes even more challenging for relational properties, as the user needs to concomitantly check the values of program variables across multiple (i.e., more than one) program runs.
Nonetheless, this is a very important step which the user needs to perform in order to improve the analyzed specification and/or code.
The process of verifying software is an iterative one, as described in~\cite{hvc17} as follows:
``Until the verification succeeds,
\begin{enumerate*}[label={(\alph*)}]
\item\label{it:failedAttempt} failed attempts have to be inspected in order to understand the cause of failure and
\item the next step in the proof process has to be chosen.''
\end{enumerate*}

\paragraph{Contribution.} The contribution of this paper is \emph{DIbugger}, a novel tool that supports the user in understanding the reason why some input leads to a violation of a \emph{k}-relational property.
This input may be provided by a verification tool as counterexample for such a property.
Thereby, we enhance step~\ref{it:failedAttempt} in the iterative software verification process mentioned above.
DIbugger extends familiar concepts from software debugging in order to support the user in finding the points of execution of the analyzed programs which introduce a violation of the relational property.
DIbugger allows for conditional expressions and watch expressions which use conditions and expressions which refer to any or all of the analyzed programs.
Moreover, additional user assistance is provided by backwards debugging and adaptable step sizes for each analyzed program.
To the best of our knowledge, DIbugger is the first tool that addresses the problem of debugging relational properties.

\paragraph{Structure of the paper.} In \cref{sec:tool}, we present DIbugger and its main functionalities.
Furthermore, \cref{sec:examples} shows how DIbugger can be used which is illustrated by applying it on an example.
We explain the range of supported properties and present related work in \cref{sec:related} and finally conclude in \cref{sec:conclusion}.

%% file: tool.tex
\section{DIbugger}
\label{sec:tool}

DIbugger\footnote{DIbugger is available at \url{https://git.scc.kit.edu/py8074/dibugger}} is a relational debugger for the \emph{WLANG} programming language.
WLANG is a subset of the C programming language and supports sequential, interprocedural programs.
Dynamic memory allocation and object-oriented programming features are not yet supported.

\begin{figure}[h]
\begin{center}
\includegraphics[scale=0.35]{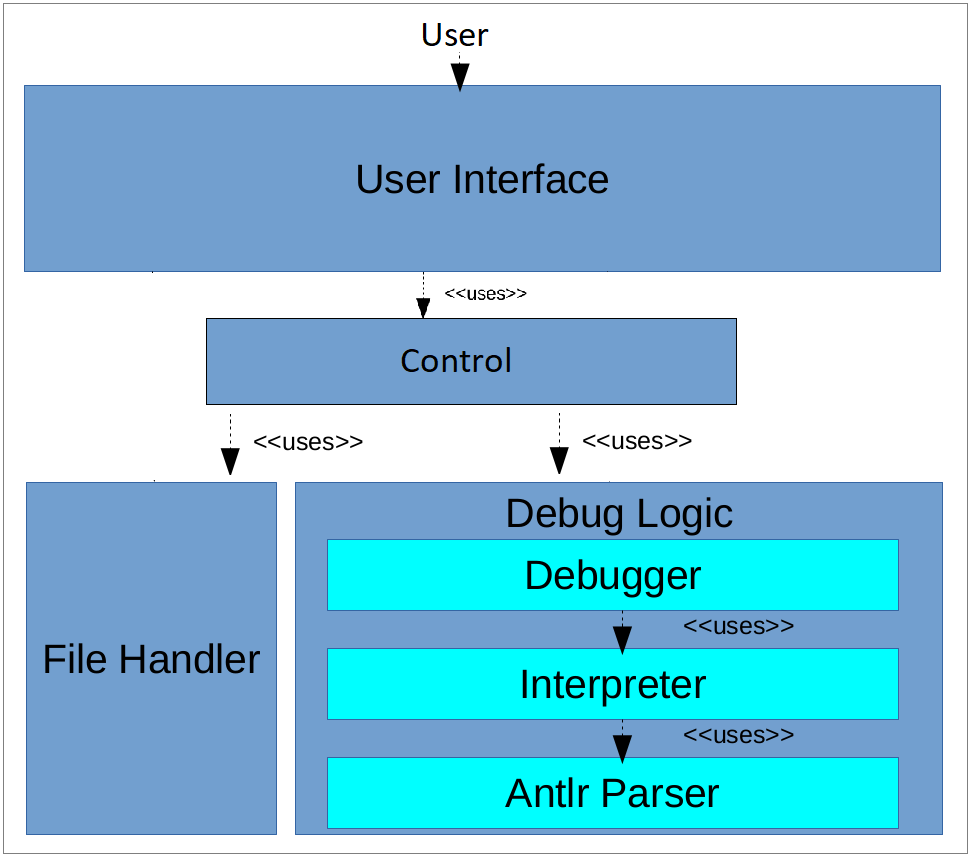}
\caption{The architecture of DIbugger}
\label{fig:architecture}
\end{center}
\end{figure}

As shown in \cref{fig:architecture}, DIbugger consists of four components which are responsible for the user interface, control, file handling and debugging respectively.
The debugging functionality is built on top of an interpreter for WLANG.
The interpreter generates the trace (i.e., the sequence of values of a program's variables at each point of its execution) of each analyzed program from the given inputs.
The debugger works on those traces and executes the debugging operations as selected by the user.
The graphical user interface (GUI) of DIbugger is shown in \cref{fig:ui}, and the available features are explained in the following.

\subsection{Debugging Operations}

\begin{figure}
\begin{center}
\includegraphics[scale=0.35]{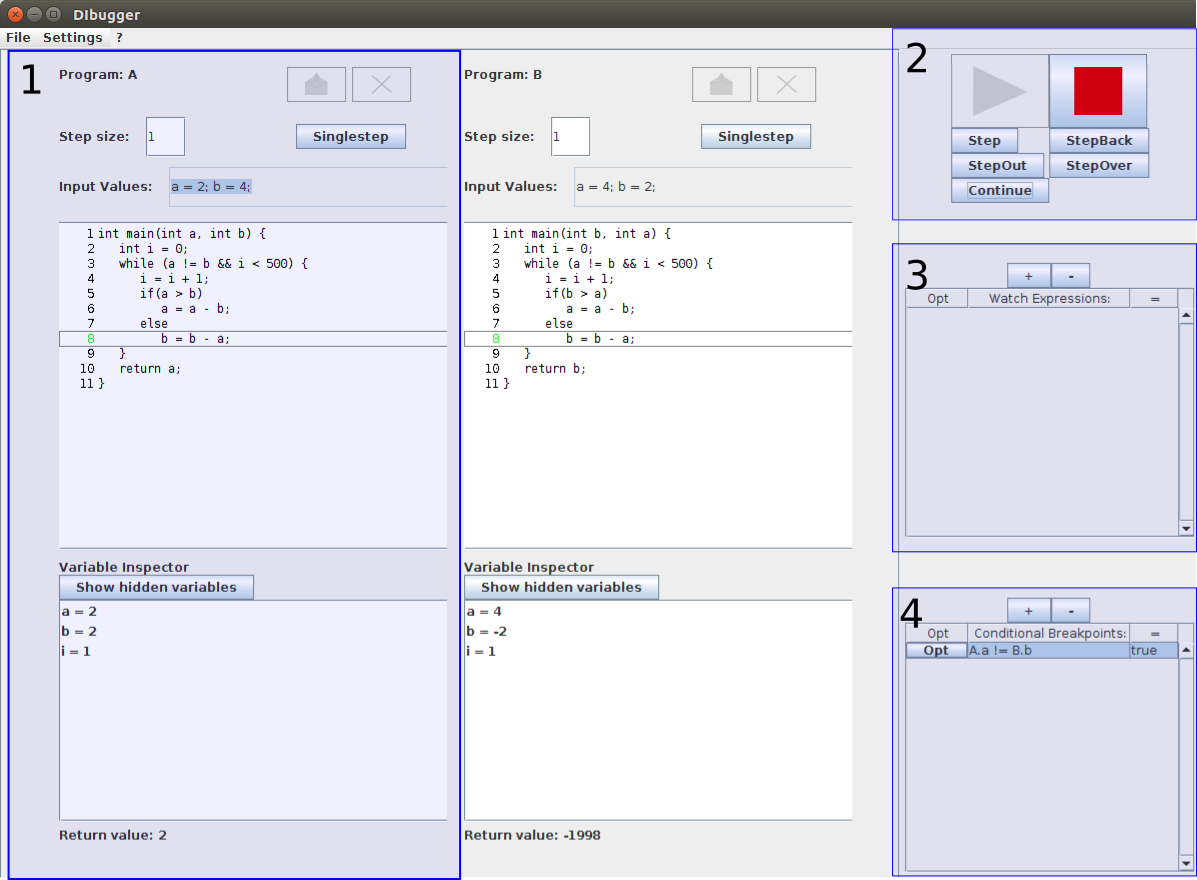}
\caption{The user interface of DIbugger}
\label{fig:ui}
\end{center}
\end{figure}

The buttons for the debugging operations are situated in the top right part of the GUI (see \cref{fig:ui}, in the highlighted~area~\textbf{2}).
The \emph{Play} and \emph{Stop} buttons allow for switching between the debug and the edit mode, respectively.
In the edit mode, the analyzed programs can be modified, and once the user switches into debug mode, the traces are generated and the programs can be debugged.
Moreover, as seen below the \emph{Play} and the \emph{Stop} button, the following buttons provide the specific debugging functionalities:

\begin{itemize}
\item \emph{Step}: The execution of each analyzed program advances by the \emph{step size}, i.e., the amount of execution points, specified by the user in the program panel (highlighted~area~\textbf{1} described in \cref{sec:progpanel}).
Thereupon, the variable values computed by the instructions in these traversed execution points can be inspected for each program in the \emph{variable inspector} in the program panel.
The user can use different step sizes for the analyzed programs in order to both keep the program executions synchronized and examine loops where the numbers of instructions vary between the programs.
Depending on the analyzed property or programs, mutable step sizes allow the user to keep the programs in lockstep even when some programs progress faster than others.

\item \emph{StepBack}: The execution of each analyzed program moves one step back.

\item \emph{StepOut}: The execution of each analyzed program either jumps out of the respective current method or---if already in the outermost method---moves to the end of the main method.

\item \emph{StepOver}: The execution of each analyzed program performs a normal step, but does not step into any traversed method call.

\item \emph{Continue}: The execution of each analyzed program advances to the next---whichever comes first---breakpoint, conditional breakpoint evaluating to true (see \cref{sec:watchexpr}), or end of the main method.
\end{itemize}

\subsection{Program Panels}
\label{sec:progpanel}

The central part of the GUI contains one \emph{program panel} for each analyzed program, as illustrated in the highlighted~area~\textbf{1} of \cref{fig:ui}.
The two buttons at the top of the program panel allow the user to add a new program panel or to remove the existing panel.
Below these buttons, the user can modify the step size for debugging, or, by using the \emph{singlestep} button, perform a single debugging step only in the selected program.
Further down, the user must provide the debugging inputs for the respective program.
In the center of the program panel, the user can inspect the analyzed program and set breakpoints anywhere within the code.
These breakpoints correspond to synchronization points.
When pressing the \emph{Continue} button, all analyzed programs advance to their next breakpoint or---when activated before---conditional breakpoint.
Below the analyzed program, the variable inspector shows the current values for all program variables in the scope of the current execution point, as well as the return value of the main method.
Note that each program panel has a unique identifier (e.g., the highlighted program panel in \cref{fig:ui} has identifier \emph{A}) which the user must use when writing watch expressions and conditional breakpoints (see \cref{sec:watchexpr}) referring to the program's variables.

\subsection{Watch Expressions and Conditional Breakpoints}
\label{sec:watchexpr}

When debugging relational properties, the user needs to constantly compare the values of the variables in all analyzed programs.
In the context of relational verification, she would need to check whether certain \emph{relational invariants} hold at some points of interest.
However, repeating this in every step of the debugging process would be a tiresome task.
In order to reduce her effort, she can insert \emph{watch expressions} and \emph{conditional breakpoints} within the highlighted areas \textbf{3} and \textbf{4} in \cref{fig:ui}, respectively.

Watch expressions are WLANG expressions which may contain variable identifiers from any of the analyzed programs.
They help the user to compare values between the execution points of the analyzed programs.
At each point of the debugging process, the value of the expression is computed and the result is displayed.
This feature allows the user to check at any time whether certain relational invariants hold.

Conditional breakpoints are boolean expressions which are evaluated at every execution point reached with the step sizes specified by the user.
They help the user to find the execution points of her interest.
If the expression evaluates to true, then the execution of the analyzed programs halts at that execution point.
Conditional breakpoints allow the user to search for execution points in which relational invariants are violated.
Using the \emph{Opt} button for both watch expressions and conditional breakpoints allows for setting a program scope.
A program scope consists of two line numbers for specifying start and end of the program segment in which the variables in the WLANG expressions are to be evaluated.
Thus, if the program execution is outside of the specified scope, the value of the watch expression is unknown and, for conditional breakpoints, the execution does not halt outside of the scope.

%% file: examples.tex
\section{Using DIBugger on an Example}
\label{sec:examples}

In the following, we illustrate how DIbugger can be used by applying it on an example for debugging program equivalence in the scope of software evolution.
We use two programs, an implementation of Euclid's algorithm for computing the greatest common divisor as shown in \cref{prog1}, and a modified implementation of Euclid's algorithm as shown in \cref{prog2}.
The user modifies the original implementation based on the assumption that $\mathit{gcd}(a,b) = \mathit{gcd}(b,a)$ holds.
She reversed the condition in the \texttt{if}-statement in line~$5$ such that in line~$10$ the method returns the value of the variable \texttt{b} instead of the value of the variable \texttt{a}.

\begin{minipage}[t]{.40\textwidth}
\begin{lstlisting}[label=prog1, caption=Correct Euclid's Algorithm,captionpos=b,numbers=left]
int main(int a, int b) {
   int i = 0;
   while (a != b && i < 500) {
      i = i + 1;
      if (a > b)
         a = a - b;
      else
         b = b - a;
   }
   return a;
}
\end{lstlisting}
\end{minipage}\begin{minipage}[t]{.175\textwidth}\hfill\end{minipage}\begin{minipage}[t]{.40\textwidth}
\begin{lstlisting}[captionpos=b,label=prog2, caption=Incorrect Euclid's Algorithm]
int main(int b, int a) {
   int i = 0;
   while (a != b && i < 500) {
      i = i + 1;
      if (b > a)
         a = a - b;
      else
         b = b - a;
   }
   return b;
}
\end{lstlisting}
\end{minipage}

In order to check whether the implementation behaves identically to the original one, a relational verification tool may be used.
In our example, the relational verification tool returns a counterexample that consists of an identical input for both programs, e.g., $(a=2, b=4)$.
With this input, the user can execute the two programs and observe that the original implementation returns $2$ and the modified implementation returns $-1998$.
While this shows that the two implementations behave differently with respect to the result value, it does not help the user in understanding \emph{why} this is the case.
Therefore, the user needs to debug the two programs in parallel.

When the user starts the debugging process, the return value of each of the two programs is shown below the respective program.
This is possible as the interpreter of DIbugger first executes the analyzed programs and then generates their traces.
Note that within our application scenario, these computations may not cause any relevant performance problems.
Beyond DIbugger's application for understanding counterexamples, the real bottleneck regarding the programs' sizes and complexities lies within the preceding verification task performed by the verification tool.

Generating the trace for each program allows debugging features such as stepping backwards, which is very useful when debugging multiple programs side by side.
In case the user performs too many debugging steps and advances beyond the execution point of her interest, she can simply step back in the programs until she reaches her point of interest.
With two conventional debuggers---however---she would be required to restart the debugging process.
Furthermore, the precomputed traces enable the support of conditional breakpoints, where the user can find pairs of execution points which violate a relational invariant.

In our example, the user can specify that the relational invariant $A.a == B.b$ needs to hold after every debugging step.
With this invariant as a conditional breakpoint, pressing the \emph{Continue} button stops the execution in both programs at line $8$ with the values $2$ and $-2$ for $A.a$ and $B.b$ respectively.
The user then examines these values and understands that she forgot to switch the \emph{then}- and \emph{else}-case in the \emph{if}-statement.
Afterwards, she enters the editing mode, edits the second program by applying the necessary changes, and then enters the debugging mode again.
Finally, both programs return the (same) value $2$.

We see already in this simple example that the analysis using conventional debuggers, which only allows to inspect a single program execution simultaneously, would be more difficult.
As in conventional debuggers the user must guide the debugging process for all programs separately, she cannot use watch expressions and conditional breakpoints for finding tuples of execution points which violate the relational property.

%% file: related.tex
\section{Supported Properties and Related Work}
\label{sec:related}

In the following, we elaborate on two relevant aspects closely related to DIbugger.
First, we illustrate the range of (relational) properties which DIbugger supports additionally to the (functional) properties which are also supported by conventional debugging tools.
Second, we cover approaches related to DIbugger in the sense that debugging is performed in the scope of or combined with the task of formal verification.

\paragraph{Supported properties.} DIbugger supports the inspection of counterexamples for any \emph{k-safety property}, i.e., properties that can be refuted by at most $k$ traces~\cite{ClarksonS10}.
A prominent target for verification of $k$-safety properties is program equivalence.
Verification approaches for program equivalence exist, e.g., for C~programs~\cite{KieferKlebanovUlbrich2017} or PLC~software~\cite{BeckertUVW15}, and allow both verification and counterexample generation.
Another example of relational properties are information flow properties which target the problem whether certain outputs can be influenced by certain inputs of the program.
The \KeY{} theorem prover~\cite{key} supports such properties~\cite{SchebenGreiner2016} and can also generate counterexamples.
A great variety of relational properties exists, e.g., when specifying fairness properties in the context of social choice theory.
Therein, a prominent example are voting algorithms which take the individual votes and compute the elected candidates, with relational properties such as monotonicity, anonymity, neutrality or reinforcement.
Relational properties for voting algorithms can also being verified using formal methods~\cite{BeckertBormerKirstenEtAl2016} and the generated counterexamples to such properties can greatly enhance the understanding and selection of such algorithms~\cite{KirstenCailloux2018}.

\paragraph{Related work.} Debugging itself is a well-known and established technique from software engineering and implemented in a multitude of software development environments.
However, we did not find any work on the process of simultaneously debugging multiple programs in a synchronized or relational fashion.
One related idea is the concept of \emph{delta debugging}, which searches for failure causes, i.e., how and when the infection causing the software defect has been propagated~\cite{cleveZ05}.
Cleve and Zeller attempt to obtain the smallest possible subset of relevant variables by performing a search over both the chain of applied changes and the original variables which might have caused the infection.
Thereby, infectious state differences are automatically narrowed down both in time and in space while requiring logarithmic to quadratic runtime.

Another approach more oriented towards understanding counterexamples is the \texttt{explain} tool, which works interactively~\cite{groceKL04}.
Groce et al.\ perform a causal slicing algorithm based on bounded model checking by first producing a counterexample and then computing a successful execution most similar to the failing run using distance metrics.
Guided by this distance metrics, the user searches the cause which seems most convincing to her, and finally uses the bounded model checker to verify that the suspected cause is indeed a valid explanation for the failing run.
Comparing multiple programs is also interesting for inspecting concurrent programs.
Jalbert and Sen apply a greedy slicing technique to simplify complex buggy traces from concurrent program executions to gain a better understanding for the cause of the failure~\cite{jalbertS10}.

Moreover, ideas from debugging have also been applied to deductive program verification in order to inspect proofs for program correctness based on logical calculi.
Exploiting the technique of symbolic execution, Hentschel et al.\ devised a symbolic execution debugger which symbolically analyses all possible program states based on the program's formal precondition~\cite{hentschelHB2016}.
This technique allows the inspection of failed proof attempts which help understanding possibly undesired program behaviour.
Finally, the debugging mindset can also directly integrated in full program verification on the basis of a compact proof language.
Beckert et al.\ have instrumented the\KeY{} theorem prover for Java programs in order to perform interactive what-if-analyses in a user-friendly fashion directly on the proof object~\cite{hvc17}.
The user can directly manipulate on the proof using a kind-of proof meta language also allowing to experiment by coming up with and trying out her own assumptions for gaining a detailed understanding on why the proof did not succeed (yet).

%% file: conclusion.tex
\section{Conclusion and Future Work}
\label{sec:conclusion}

We presented DIbugger, a tool that helps the user understand why the verification of a relational property failed.
The tool can be used as a counterexample analyzer for many verification approaches in various scenarios and use cases ranging from regression verification of safety critical system to the verification of information flow properties or the verification of social choice properties.

Moreover, we plan to extend the supported language features to heap-based data structures, and support the automatic suggestion of useful conditional breakpoints or watch expressions, depending on the analyzed relational property.
Further ideas to go from here are to enrich DIbugger by property-specific breakpoints in order to better-support specific use cases, or to extend the current breakpoints and watch expressions to quantitative program comparisons.
Such ideas could also be integrated in a larger framework guided by counterexamples for abstracting the program.
Finally, we would like to apply DIbugger to larger use cases to gain more experiences on its scalability and usability for specific use cases.